# First-principles studies of the structural, electronic, and optical properties of a novel thorium compound $Rb_2Th_7Se_{15}$

**M.G. Brik**[1]

*Institute of Physics, University of Tartu, Riia 142, Tartu 51014, Estonia*

**Abstract**

The structural, electronic, and optical properties of a recently synthesized thorium compound $Rb_2Th_7Se_{15}$ have been calculated in the density functional theory framework for the first time. The calculated direct band gap was 1.471 eV (generalized gradient approximation) and 1.171 eV (local density approximation), with both results being close to the experimental result of 1.83 eV. High covalency/iconicity of the Th-Se/Rb-Se bonds was demonstrated by calculating effective Mulliken charges of all ions. The polarized calculations of the complex dielectric function are presented; dependence of the calculated index of refraction was fitted to the Sellmeyer equation in the wavelength range from 500 to 2500 nm.

## 1. Introduction

Very recently [1] a new compound with the chemical composition $Rb_2Th_7Se_{15}$ has been synthesized. It was reported to be just the second ternary compound (after $RbTh_2Se_8$) in the Rb/Th/Se system. The details of its synthesis, crystal structure and optical band gap measurements were all presented in Ref. [1]. It should be mentioned that thorium is a radioactive element, and because of this represents a serious danger for health. This can be one of the most important reasons why the properties of the thorium-bearing compounds remain to a large extent unexplored. Several groups of other ternary compounds containing thorium and differing by the composition have been mentioned in Refs. [1, 2].

In the present paper, the next step towards further and deeper characterization of the $Rb_2Th_7Se_{15}$ material is made by performing detailed bench-mark first-principles calculations of its structural, electronic, and optical properties. Optimization of the crystal structure at the ambient pressure yielded an excellent agreement with the experimental structural data. Composition of the electronic bands has been interpreted in terms of the atom-projected density of states (DOS), which also allowed for an analysis and interpretation of the calculated optical absorption spectra. Particular attention has been paid to the bonding properties and distribution of the electron density difference in the space between the crystal lattice ions. The first estimations of the refractive index of this material and its dependence on the wavelength fitted to the Sellmeyer equation are also reported. It is believed that such a theoretical work, if taken together with the experimental research from Ref. [1], provides the readers with additional information on the properties of $Rb_2Th_7Se_{15}$.

---

[1] Corresponding author. E-mail: brik@fi.tartu.ee Phone +372 7374751

## 2. Crystal structure and details of calculations

$Rb_2Th_7Se_{15}$ crystallizes in the monoclinic space group $P2_1c$ (No. 14) with four formula units in one unit cell. The formal charges of the ions, as implied by their valences and chemical formula, are +1 for Rb, +4 for Th and -2 for Se. Table 1 collects the main structural information on this compound, whereas the overall appearance of the crystal structure is depicted in Fig. 1.

Following Ref. [1], there are two inequivalent Rb positions, seven inequivalent Th positions and fifteen inequivalent Se positions, which makes the structure be rather complicated. The coordination numbers of both Rb and Th ions vary from 7 to 9, depending on the crystallographic position, and the local symmetry of each site is very low. The Rb ions positions can be described as those located within the three-dimensional channels formed by the selenium polyhedra around thorium ions (Fig. 1).

All reported here first-principles calculations were performed using the CASTEP module [3] of Materials Studio. The generalized gradient approximation (GGA) with the Perdew-Burke-Ernzerhof [4] and the local density approximation (LDA) with the Ceperley-Alder-Perdew-Zunger (CA-PZ) functional [5,6] were used to treat the exchange-correlation effects. The plane-wave basis set cut off energy was set at 410 eV; the Monkhorst-Pack k-points mesh was 1×2×1 for the structural optimization (which is justified by large values of the lattice constants) and 2×3×1 for the calculations of the optical properties and density of states (DOS); the electronic configurations were $4s^24p^65s^1$ for Rb, $6s^26p^66d^27s^2$ for Th, and $4s^24p^4$ for Se. The convergence criteria were as follows: energy $10^{-5}$ eV/atom, force 0.03 eV/Å, stress 0.05 GPa, and atomic displacements $10^{-3}$ Å.

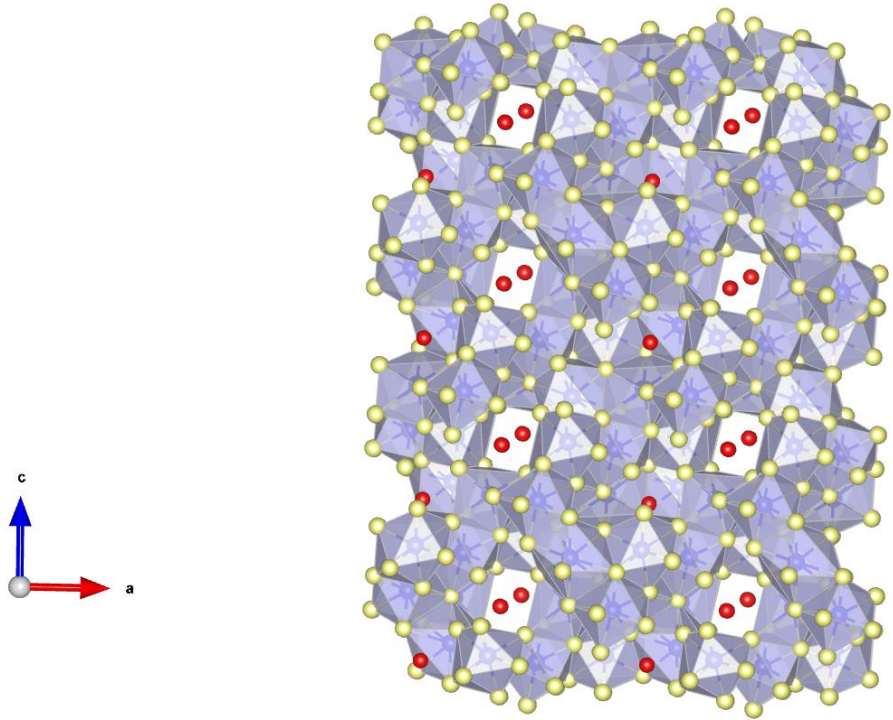

Fig. 1. Overall view (along the *b* axis) of the $Rb_2Th_7Se_{15}$ crystal structure. Coordination polyhedra around thorium ions are shown. The rubidium ions are shown as unbounded between the polyhedral. Drawn with VESTA [7].

## 3. The results of calculations: structural and electronic properties

Table 1 offers a direct comparison of the experimental [1] and calculated in the present work crystal data for $Rb_2Th_7Se_{15}$. As seen from the Table, agreement between the structural parameters (experimental and optimized) is excellent in the case of the GGA calculations (with the largest deviation between the experimental and calculated lattice parameters being 0.4% only for the *b* constant). In the case of the LDA results the maximum deviation was found to be about 1.8 % for the *c* constant.

Table 1. Comparison between the experimental and calculated structural data and density for $Rb_2Th_7Se_{15}$.

| | Experiment [1] | Calculated (this work) | |
| --- | --- | --- | --- |
| | | GGA | LDA |
| $a$, Å | 15.2384 | 15.26447 | 14.99623 |
| $b$, Å | 7.8625 | 7.89302 | 7.76689 |
| $c$, Å | 21.7614 | 21.75446 | 21.37209 |
| $\beta$, ° | 90.534 | 90.3277 | 90.7578 |
| $V$, Å$^3$ | 2607.2 | 2621.00 | 2489.08 |
| $\rho$, kg/m$^3$ | 7591 | 7551 | 7952 |

It can be also noted that the GGA-calculated density is practically coinciding with the experimental estimation, differing just by about 0.5%. The fractional coordinates of all atoms in a unit cell of $Rb_2Th_7Se_{15}$ are all collected in Table 2; again, a very good agreement between the experimental and calculated data can be noticed.

Additional information related to the geometry of the Rb and Th positions is presented in a separate supplementary file, which shows the coordination of all ions and all calculated Rb-Se and Th-Se distances.

The calculated band structure of $Rb_2Th_7Se_{15}$ (in the region of the band gap only) is shown in Fig. 2a. Fig. 2b visualizes the Brillouin zone and the path along which the energy band structure is shown in Fig. 2a. The calculated band gap was equal to 1.471 eV (GGA) and 1.171 eV (LDA), which are both underestimated if compared to the experimental result of 1.83 eV [1]. This is a common feature of the density functional theory-based calculations. It can be overcome by introducing an *a posteriori* correction called scissor factor, which just shifts up the conduction band. In our case, such a correction was 0.359 eV (GGA) and 0.659 eV (LDA).

As was mentioned in Ref. [1], $Rb_2Th_7Se_{15}$ is characterized by an indirect band gap. However, such a conclusion is not supported by the present calculations, since the top of the valence band and the bottom of the conduction band are realized at the same point − the Brillouin's zone center G, thus placing the studied compound into a group of the direct-band gap semiconductors. More detailed experimental studies might be needed to clarify such an issue. It should be kept in mind that the difference between the top of the valence band at the G point and the energies of the valence band in the B − D segment is only about 0.1 eV, which may contribute to some difficulties in distinguishing the experimental result regarding the nature of the band gap (direct or indirect).

Table 2. Comparison between the experimental and calculated atomic positions (in the units of the lattice constants from Table 1) for $Rb_2Th_7Se_{15}$. The last column shows the effective Mulliken charges in units of the proton charge.

|      | Experiment [1] | | | GGA | | | LDA | | | Mulliken charge, GGA/LDA |
|------|----------|---------|---------|---------|---------|---------|---------|---------|---------|-------------|
|      | x | y | z | x | y | z | x | y | z | |
| Th1  | 0.096135 | 0.41713 | 0.422871 | 0.09285 | 0.41891 | 0.42032 | 0.09563 | 0.41780 | 0.42096 | +0.14/-0.07 |
| Th2  | 0.117361 | 0.52457 | 0.060927 | 0.11747 | 0.52568 | 0.06007 | 0.11686 | 0.52315 | 0.05996 | +0.20/+0.03 |
| Th3  | 0.212513 | 0.02613 | 0.104890 | 0.21185 | 0.02600 | 0.10651 | 0.21326 | 0.02505 | 0.10556 | +0.13/-0.06 |
| Th4  | 0.248592 | 0.56904 | 0.244714 | 0.25035 | 0.57050 | 0.24268 | 0.24648 | 0.57379 | 0.24345 | +0.11/-0.12 |
| Th5  | 0.421263 | 0.42270 | 0.085704 | 0.42157 | 0.42055 | 0.08601 | 0.42340 | 0.42099 | 0.08594 | +0.15/-0.06 |
| Th6  | 0.546627 | 0.47570 | 0.274196 | 0.54561 | 0.47599 | 0.27373 | 0.54481 | 0.47697 | 0.27453 | +0.16/+0.01 |
| Th7  | 0.737239 | 0.44536 | 0.114202 | 0.74068 | 0.44785 | 0.11316 | 0.73489 | 0.44972 | 0.11403 | +0.13/-0.08 |
| Rb1  | 0.04496  | 0.05047 | 0.26820  | 0.04473 | 0.05276 | 0.26770 | 0.04506 | 0.05188 | 0.26785 | +0.41/+0.21 |
| Rb2  | 0.39228  | 0.45216 | 0.44191  | 0.39228 | 0.45281 | 0.44110 | 0.39049 | 0.45399 | 0.44137 | +0.52/+0.29 |
| Se1  | 0.04612  | 0.16431 | 0.04712  | 0.04671 | 0.16517 | 0.04790 | 0.04563 | 0.16364 | 0.04668 | -0.29/-0.21 |
| Se2  | 0.06744  | 0.80868 | 0.14419  | 0.06866 | 0.80788 | 0.14397 | 0.06950 | 0.80584 | 0.14605 | -0.09/+0.08 |
| Se3  | 0.07827  | 0.03577 | 0.42810  | 0.07491 | 0.03766 | 0.42988 | 0.07847 | 0.03564 | 0.42656 | -0.08/+0.01 |
| Se4  | 0.12742  | 0.69728 | 0.33265  | 0.12990 | 0.70284 | 0.33316 | 0.12680 | 0.70212 | 0.33321 | -0.12/0.00 |
| Se5  | 0.14100  | 0.32883 | 0.17457  | 0.13944 | 0.33335 | 0.17366 | 0.13981 | 0.32817 | 0.17433 | -0.03/+0.10 |
| Se6  | 0.23211  | 0.29274 | 0.34331  | 0.23082 | 0.29487 | 0.34088 | 0.23222 | 0.29524 | 0.34077 | -0.05/+0.07 |
| Se7  | 0.25986  | 0.31109 | 0.01081  | 0.25924 | 0.31493 | 0.00768 | 0.26049 | 0.31124 | 0.00948 | -0.19/0.00 |
| Se8  | 0.28433  | 0.66654 | 0.11859  | 0.28669 | 0.66642 | 0.11787 | 0.28645 | 0.66665 | 0.11718 | -0.28/-0.23 |
| Se9  | 0.38653  | 0.33413 | 0.21431  | 0.38821 | 0.33415 | 0.21509 | 0.38445 | 0.33497 | 0.21461 | -0.22/-0.16 |
| Se10 | 0.40359  | 0.04847 | 0.09532  | 0.39889 | 0.05490 | 0.09697 | 0.40479 | 0.04553 | 0.09481 | -0.07/+0.06 |
| Se11 | 0.46900  | 0.16314 | 0.34294  | 0.47231 | 0.16597 | 0.34084 | 0.46427 | 0.16147 | 0.34411 | -0.13/-0.02 |
| Se12 | 0.58827  | 0.36868 | 0.02811  | 0.59130 | 0.37151 | 0.02826 | 0.58850 | 0.36281 | 0.02617 | -0.12/-0.01 |
| Se13 | 0.60250  | 0.20160 | 0.18101  | 0.60213 | 0.20387 | 0.18076 | 0.60202 | 0.20441 | 0.18107 | -0.12/+0.08 |
| Se14 | 0.73852  | 0.45303 | 0.26021  | 0.73760 | 0.45020 | 0.26075 | 0.73813 | 0.45823 | 0.26001 | -0.10/+0.02 |
| Se15 | 0.79877  | 0.17684 | 0.01244  | 0.80131 | 0.17435 | 0.01206 | 0.79865 | 0.17794 | 0.01307 | -0.05/+0.07 |

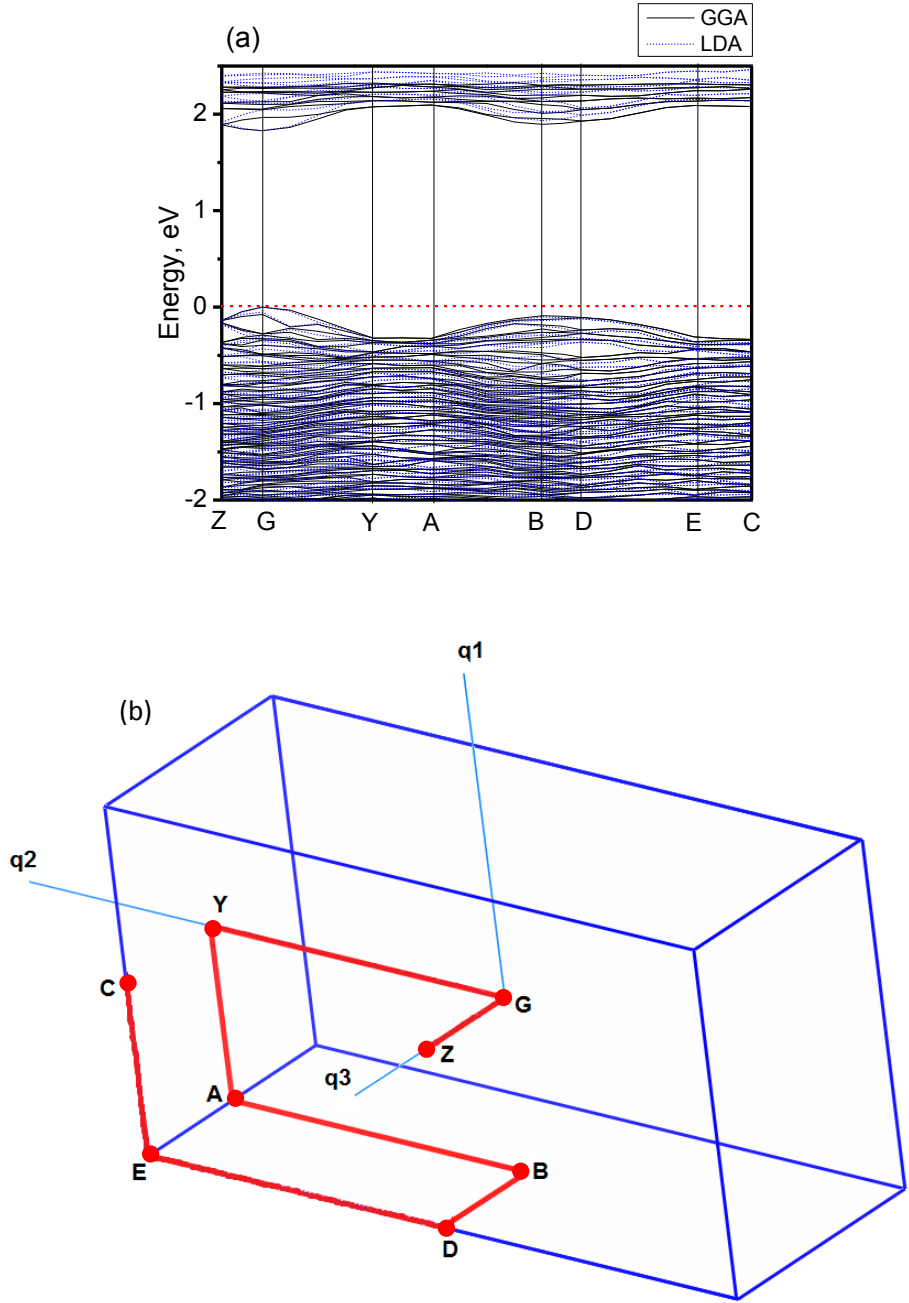

Fig. 2. Calculated band structure of $Rb_2Th_7Se_{15}$ (a) and Brillouin zone with a path along which the band structure is shown (b).

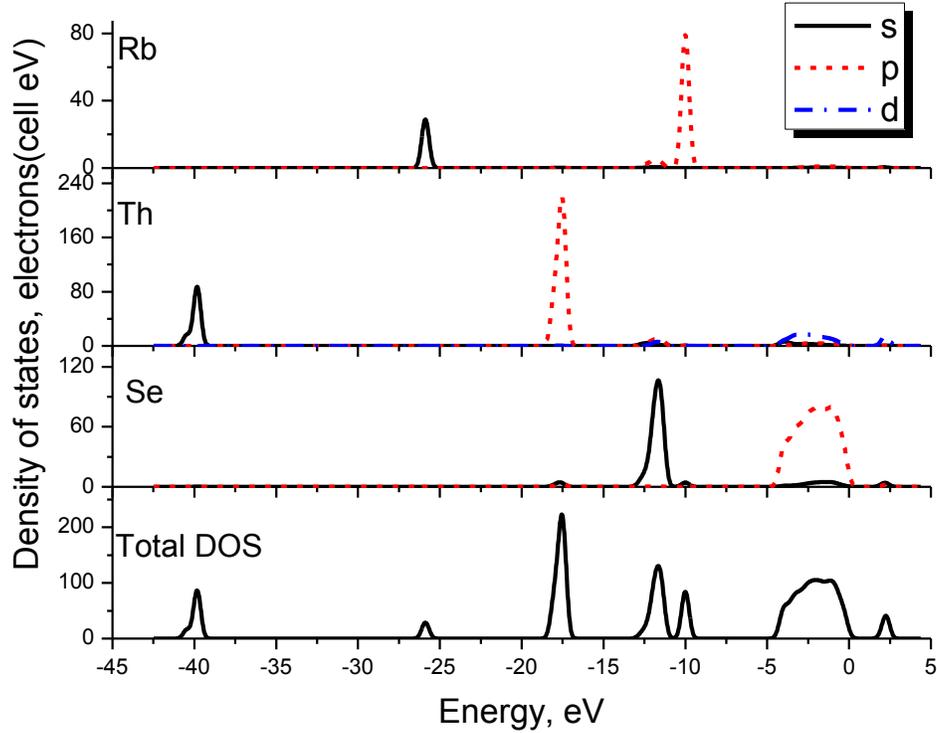

Fig. 3. Calculated DOS diagrams for $Rb_2Th_7Se_{15}$. From top to bottom: Rb, Th, Se partial DOS and total DOS.

With the help of the DOS diagrams shown in Fig. 3, an assignment of the electronic bands is straightforward. A very narrow conduction band between approximately 1.8 eV and 3 eV is composed mainly of the Th 6d and Rb 5s states. A wide valence band from about -5 eV to 0 eV is made by the Se 4p states. A slight hybridization of the Se 4p and Th 6d states can be noticed due to the formation of chemical bonds between these atoms. Two narrow bands at about -10 eV and -12 eV are produced by the Rb 4p and Se 4s states, respectively. A deeper band located at about -17.5 eV is due to the Th 6p states. Finally, very deeply localized bands at -26 eV and -40 eV are produced by the Rb 4s and Th 6s states, respectively.

Table 2 also contains the calculated effective Mulliken charges [8] of all ions in $Rb_2Th_7Se_{15}$. All charges are remarkably different from those, which can be expected from the formal chemical formula, especially for Th and Se. This is a clear indication of high covalency of the Th-Se bonds. It also can be noted that the LDA gives slightly positive charges for some of the Se ions, whereas the GGA shows all Se ions to bear a negative charge, and all Rb and Th ions to have a positive charge, which (the signs, but not the values, of course) agrees with the expectations from the chemical formula. Thus, the GGA turns to be a better approximation for this particular compound.

The Rb-Se bonds are more ionic, as is also illustrated by Fig. 4, which shows the cross-sections of the electron density difference in the space surrounding Th, Se and Rb ions. Formation of the covalent bonds between the Th and Se ions is clearly seen, whereas the electron density difference distribution around Rb ions does not exhibit pronounced directional dependence.

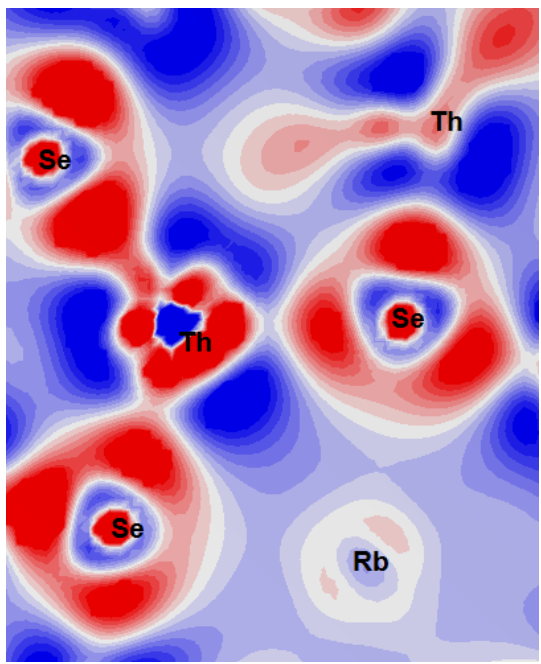

Fig. 4. Electron density difference around Rb, Th and Se in $Rb_2Th_7Se_{15}$, as calculated using GGA. The red/blue areas indicate gain/loss of electron density.

**4. The results of calculations: optical properties**

After successful optimization of the crystal structure and calculations of the electronic properties, the optical features can be calculated in CASTEP in a straightforward manner. The imaginary part $\text{Im}(\varepsilon(\omega))$ of a dielectric function $\varepsilon(\omega)$ is calculated by direct numerical evaluations of the matrix elements of the electric dipole operator between the occupied states in the valence band and empty states in the conduction band:

$$\mathrm{Im}\left(\varepsilon(\omega)\right) = \frac{2e^2\pi}{\omega\varepsilon_0} \sum_{\mathbf{k},v,c} \left|\left\langle \Psi_{\mathbf{k}}^c \left| \mathbf{u}\cdot\mathbf{r} \right| \Psi_{\mathbf{k}}^v \right\rangle\right|^2 \delta\left(E_{\mathbf{k}}^c - E_{\mathbf{k}}^v - E\right), \tag{1}$$

where $\mathbf{u}$ is the polarization vector of the incident electric field, $\mathbf{r}$ and $e$ are the electron's radius-vector and electric charge, respectively, $\Psi_{\mathbf{k}}^c, \Psi_{\mathbf{k}}^v$ are the wave functions of the conduction/valence bands, respectively, $E = \hbar\omega$ is the incident photon's energy, and $\varepsilon_0$ is the vacuum dielectric permittivity. The summation in Eq. (1) is carried out over all occupied and empty states, with their wave functions obtained in a numerical form after optimization of the crystal structure. The imaginary part $\mathrm{Im}(\varepsilon(\omega))$ is proportional to the absorption spectrum of a solid. The real part $\mathrm{Re}(\varepsilon(\omega))$ of the dielectric function $\varepsilon(\omega)$, which determines the dispersion properties and refractive index values, is estimated then by using the Kramers-Kronig relation:

$$\mathrm{Re}(\varepsilon(\omega)) = 1 + \frac{2}{\pi} \int_0^\infty \frac{\mathrm{Im}(\varepsilon(\omega'))\omega' d\omega'}{\omega'^2 - \omega^2} \tag{2}$$

The GGA- and LDA-calculated real and imaginary parts of the $Rb_2Th_7Se_{15}$ dielectric function for three different polarizations (1, 0, 0), (0, 1, 0) and (0, 0, 1) are shown in Fig. 5. A slight anisotropy of the optical properties can be noticed, especially for the imaginary part of $\varepsilon(\omega)$, where the main maximum is slightly shifted in different polarizations.

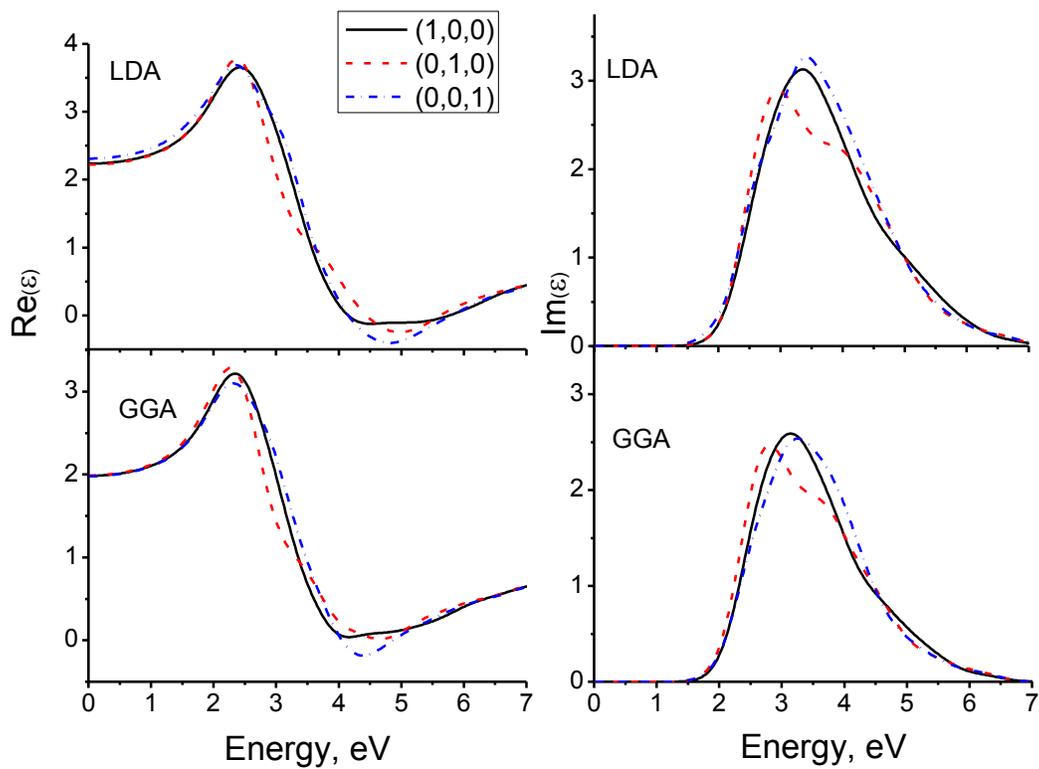

Fig. 5. Calculated real (left panel) and imaginary (right panel) parts of dielectric function for Rb$_2$Th$_7$Se$_{15}$.

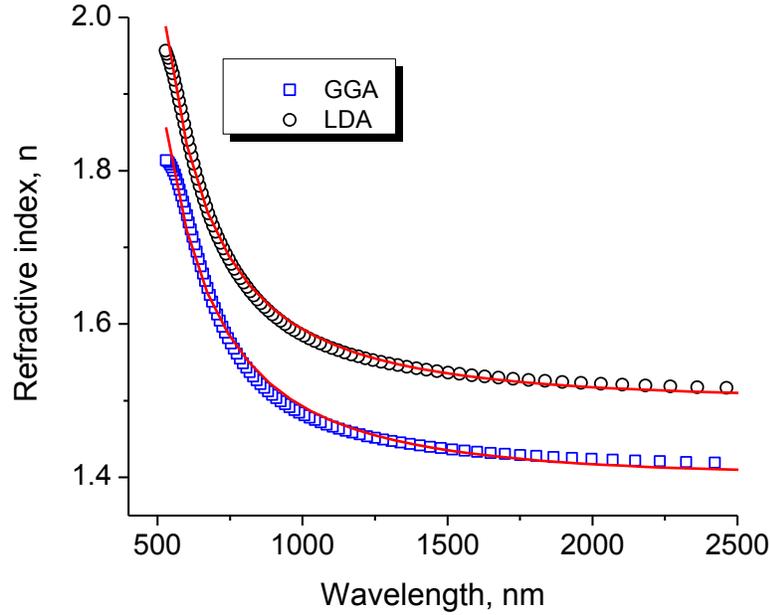

Fig. 6. Calculated values of the refractive index (symbols) and fits to the Sellmeyer equation (3) (solid lines) for $Rb_2Th_7Se_{15}$.

Additional useful information, which can be extracted from the calculated dielectric function, is the dependence of the refractive index $n$ on the wavelength $\lambda$. The Sellmeyer equation with the infrared correction [9]

$$n = A + \frac{B}{1-\left(\frac{C}{\lambda}\right)^2} - D\lambda^2 \qquad (3)$$

was used to fit the calculated dependence of $n$ on $\lambda$ (Fig. 6). As seen from Fig. 6, the quality of fit is good; the determined parameters of fit are collected in Table 3. The values of the $A$, $B$, $C$ and $D$ parameters from Table 3 allow for estimations of the value of $n$ in a wide spectral region from 500 to 2500 nm. The limiting values of $n$ when $\lambda \to \infty$ are 1.41 (GGA) and 1.51 (LDA).

Table 3. Calculated parameters of the Sellmeyer fit (Eq. (3)).

|  | $A$ | $B$ | $C$, nm | $D$, nm$^{-2}$ |
|---|---|---|---|---|
| GGA | 0.2506±0.1470 | 1.1409±0.1448 | −284.6002±13.2220 | (−5.18571±1.55679)×10$^{-10}$ |
| LDA | 0.5774±0.0609 | 0.9157±0.0595 | −312.7437±6.8596 | (−3.64192±0.09952)×10$^{-10}$ |

## 5. Conclusions

The results of the benchmark calculations of the structural, electronic, and optical properties of a recently synthesized novel ternary thorium compound $Rb_2Th_7Se_{15}$ are reported in the present paper. The optimized structural constants and fractional positions of atoms in a crystal lattice, obtained by using both GGA and LDA calculations, are in very good agreement with the experimental findings recently reported by Koscielski *et al*, especially in the case of the GGA results. Although Koscielski *et al* reported an indirect band gap in $Rb_2Th_7Se_{15}$, the calculated band structure does not confirm their conclusion and shows $Rb_2Th_7Se_{15}$ to be a direct band gap semiconductor with the band gap of 1.471 eV (GGA) and 1.171 eV (LDA), which are quite close to the experimental result of 1.83 eV. After a scissor operator of 0.359 eV (GGA) and 0.659 eV (LDA) was introduced, the density of states were calculated for all atoms in a unit cell to reveal the composition of the calculated electronic bands. The polarized calculations of the complex dielectric function for $Rb_2Th_7Se_{15}$ were performed; a slight optical anisotropy was noticed. The calculated values of the refractive index in the limit of the long wavelengths are 1.41 (GGA) and 1.51 (LDA). In addition, the wavelength dependence of the refractive index was fitted to the Sellmeyer equation in the spectral region between 500 nm and 2500 nm. The reported results are the first theoretical estimations of these properties for $Rb_2Th_7Se_{15}$.

## Acknowledgments

This study was supported by the European Union through the European Regional Development Fund (Centre of Excellence "Mesosystems: Theory and Applications", TK114), European Internationalisation Programme DoRa and Marie Curie Initial Training Network LUMINET, grant agreement No. 316906.

# First-principles studies of the structural, electronic, and optical properties of a novel thorium compound Rb$_2$Th$_7$Se$_{15}$

## Supplementary information

*M.G. Brik[2]*

*Institute of Physics, University of Tartu, Riia 142, Tartu 51014, Estonia*

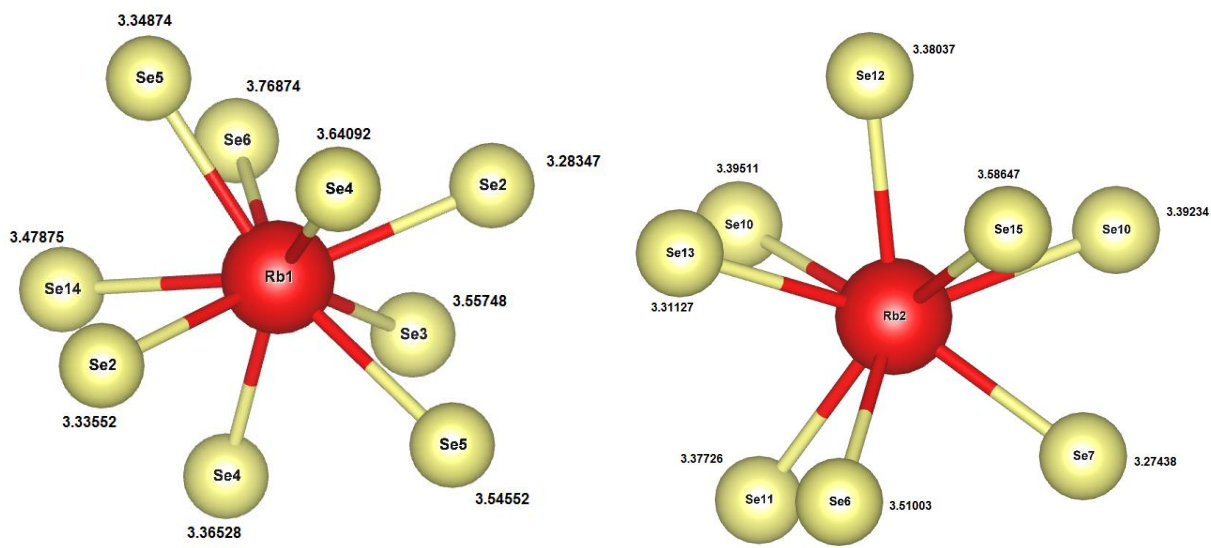

Fig. 1. Coordination of the Rb ions in Rb$_2$Th$_7$Se$_{15}$. The calculated Rb-Se distances (all in Å) are given.

---

[2] Corresponding author. E-mail: brik@fi.tartu.ee Phone +372 7374751

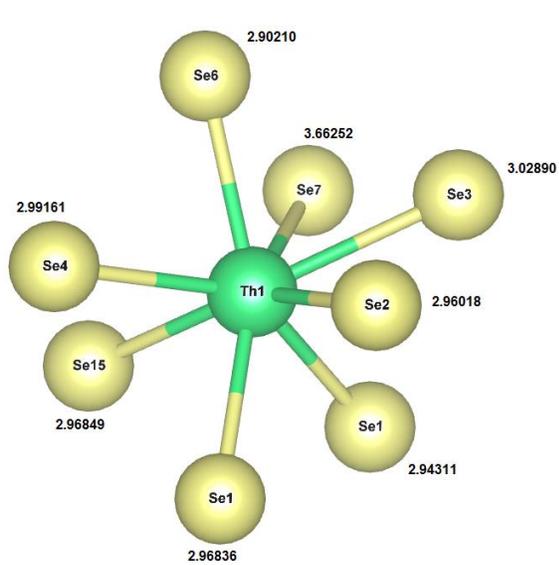
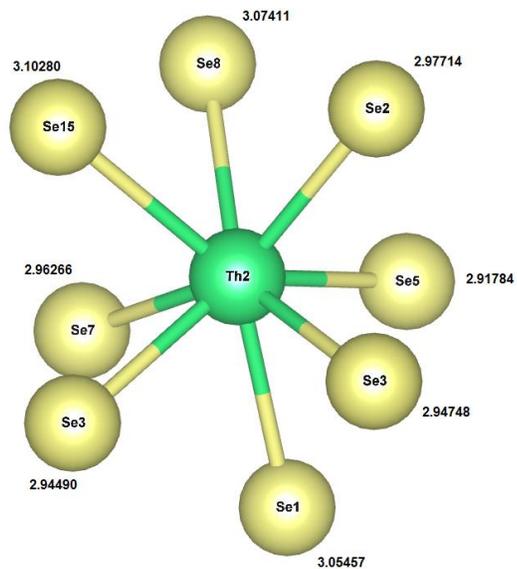
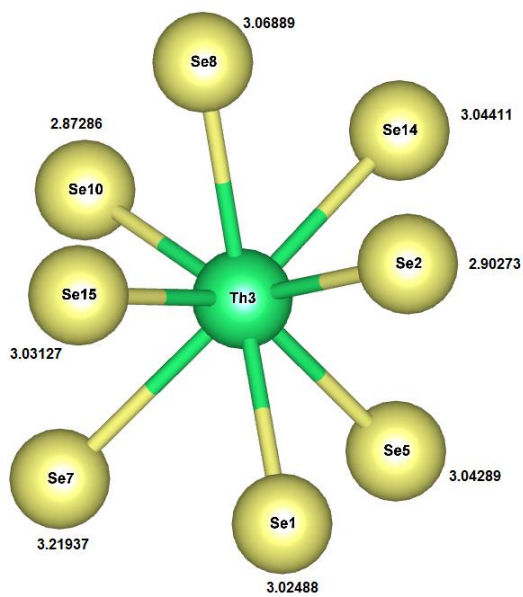
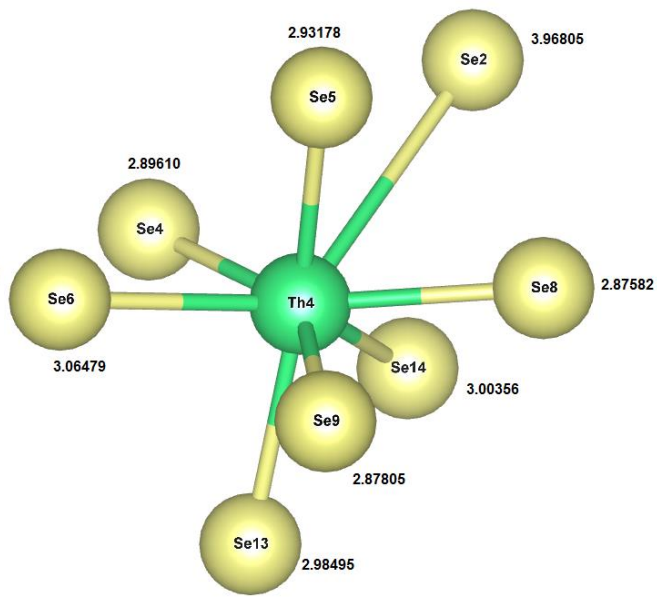

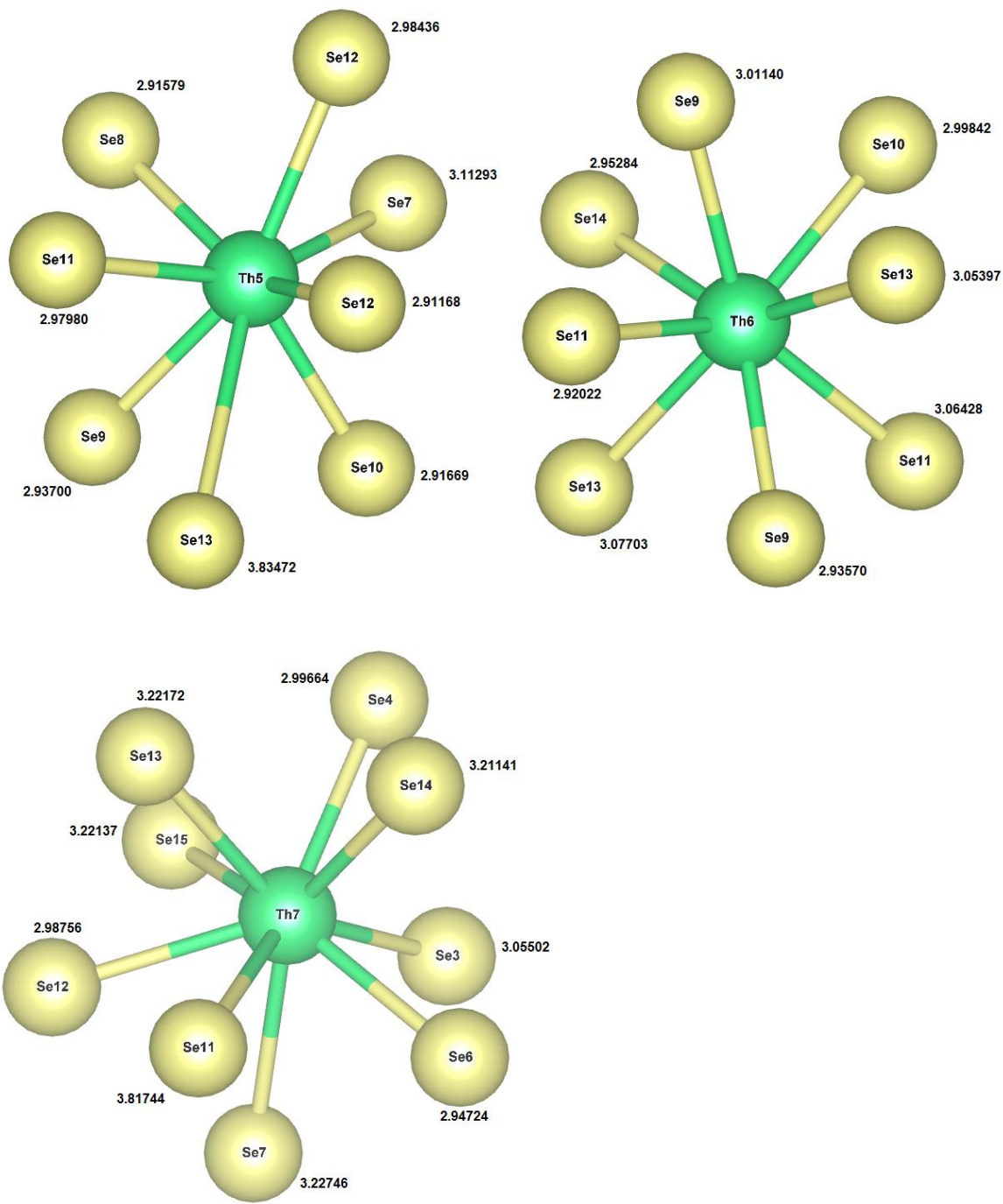

Fig. 2. Coordination of the Th ions in $Rb_2Th_7Se_{15}$. The calculated Rb-Se distances (all in Å) are given.